\documentclass[a4paper,11pt]{article}

\usepackage{geometry}
\usepackage{fancyhdr}
\usepackage{amsmath, amssymb}
\usepackage{graphicx}
\usepackage{color}
\definecolor{bg}{rgb}{0.89, 0.95, 0.71} 
\definecolor{ac}{rgb}{0.1, 0.1, 0.9} 
\definecolor{rc}{rgb}{0.77, 0.57, 0.71} 
\usepackage{listings}
\lstdefinestyle{latex}{language=TeX,
                       backgroundcolor=\color{bg},
                       basicstyle=\small\ttfamily,
                       frame=leftline,
                       xleftmargin=1.4em,
                       framexleftmargin=.8em}
\lstdefinestyle{cmdline}{
                         }
\usepackage{url}
\usepackage[pdftex,
            bookmarks,
            colorlinks=false,
            linkcolor=black,
            plainpages = false,
            pdfpagemode = UseNone,
            pdfstartview = FitH,
            citecolor = green, urlcolor = cyan, filecolor = magenta]{hyperref}

\usepackage[para]{footmisc}
\makeatletter
\long\def\@makefntext#1{\leavevmode
\@makefnmark\nobreak
\hskip.05em\relax#1%
}
\makeatother

\newcommand{\bibliolink}[2]{\href{#1}{\nolinkurl{#1}}.}

\makeatletter 

\def\section{\@startsection{section}{1}{0pt}{2.5ex plus .6ex minus
    .2ex}{1.0ex plus .15ex}{\large\bf}} 
\def\subsection{\@startsection{subsection}{2}{0pt}{1.5ex plus .3ex minus
   .1ex}{.2ex plus .1ex}{\normalsize\bf}} 
\makeatother

\newlength{\rulelength}
\makeatletter
\renewcommand{\@makecaption}[2]{%
		    \sbox{\@tempboxa}{{\small #1: #2}}
		    \ifdim \wd\@tempboxa > \hsize
		    {\small #1: #2}\par
		    \else
		    \global\@minipagefalse
		    \hbox to \hsize {\hfil {\small #1: #2}\hfil}%
		    \fi
		   \vspace{\belowcaptionskip}
		     }
\makeatother

\newcommand{\arten}[5]{\textit{#1}, #2 {\bf{#3}}, #4 (#5).}
\newcommand{\artenarx}[6]{\textit{#1}, #2 {\bf{#3}}, #4 (#5) #6.}

\begin{document}

\begin{center}

{\bf\large Simulation of propagating EAS Cherenkov radiation
over the ocean surface}\\[4mm]
Olga P. Shustova\footnote{\nolinkurl{olga.shustova@eas.sinp.msu.ru}}\\[1mm]
{\it\small Faculty of Physics, Lomonosov Moscow State University,\\Leninskie gory 1/2, Moscow 119991, Russia}\\[4mm]
Nikolai N. Kalmykov\footnote{\nolinkurl{kalm@eas.sinp.msu.ru}}, Boris A. Khrenov\\[1mm]
{\it\small Skobeltsyn Institute of Nuclear Physics, Lomonosov Moscow State University,\\Leninskie gory 1/2, Moscow 119991, Russia}
\end{center}

\begin{abstract}
We present computing results of the Cherenkov light propagation in air and water from extensive air showers developing over the ocean. Limits on zenith angles of the showers, at which the registration of f{}lashes of ref{}lected Cherenkov photons by the satellite--based detector TUS is possible, are~analyzed with consideration for waves on the ocean surface.
\end{abstract}

\section*{Introduction}%
The space experiment TUS is going to be launched in the near future for studying ultrahigh--energy cosmic rays (UHECRs)\,\cite{tus}. Its main purpose is to search for UHECRs particles over the whole celestial sphere by recording the f{}luorescence that is generated in developing of extensive air showers (EASs). This radiation is isotropic, which allows some photons to reach the satellite--based detector.

In addition, development of UHE showers is accompanied by the intensive f{}lux of Cherenkov radiation, propagating mainly along the shower axis. Nevertheless, when an EAS passes over a high--albedo surface the Cherenkov radiation can be detected at high altitudes. An example of this is the balloon experiment \cite{balloon} for recording Cherenkov photons scattered from snow--covered surfaces. According to \cite{Antonov}, the total signal of the Cherenkov radiation gives an exact estimate of the primary energy, and its spatial distribution enables to measure the shower maximum depth and therefore to estimate the mass of the primary particle.

In the case of the satellite--based detector, recording Cherenkov photons ref{}lected from the ocean surface is of great interest as their f{}lash at the end of the f{}luorescence track would allow one to determine the shower trajectory more precisely. Besides, Cherenkov photons scattered in the atmosphere produce an additional background. Calculations show that they will be superimposed on the f{}luorescence signal and blur its maximum. Thus dependence of the ``noise'' on the shower parameters should be found and subtracted from the total signal. Estimates of dif{}ferent components of the Cherenkov light, which arise in developing of showers over the ocean, are presented in \cite{Como}.

This paper presents a computing technique for propagating EAS--produced Cherenkov photons in air and water. It was used in \cite{Shustova} for estimating the ability of the TUS detector to record the ref{}lected component of the Cherenkov light. These calculations may also be useful for experimental data processing and for constructing next--generation detectors (see, for instance, the JEM-EUSO project\,\cite{jem_euso}).

\section{Calculation technique}\label{calc_tech}%
The photon propagation task can be conventionally divided into two parts: 1) generation of the photons by electrons of the shower and 2) their subsequent travel in a medium. We apply the Monte--Carlo method in both cases, with the destiny of a large number of photons (henceforth, just a photon) simultaneously predicted. Each photon is assigned by a certain weight, i.e. the number of photons ``survived" at a given calculation stage with respect to the maximum possible number of photons at the point of their generation.

We use the laboratory reference system (LRS) in which the origin $O$ is located at the intersection point of the shower axis with the plane $Oxy$ coinciding with the f{}lat ocean surface, and $Oz$--axis is directed to the top of the atmosphere. The LRS and systems to be introduced below are right--handed.

First we follow the ``scheme of photon generation'' (section \ref{sect_gener}), where the energy and direction of the parent electron, the generation depth, direction and weight of the photon are defined. Thus we obtain characteristics of the photon at the generation point, which are further used as input parameters in simulating its travel. The ``scheme of photon propagation in a medium'' (section \ref{sect_prop}) presents a sequence of scattering events in which we def{}ine the path length of the photon to the next scattering point, its direction and weight. Calculations continue until the photon reaches a given level (for instance, the height of the satellite) or its weight becomes less than a certain preset value. Ref{}lection and refraction of the photons at the ``air--water'' interface are considered with allowance for waves on the ocean surface (section \ref{sect_waves}).

\section{Scheme of photon generation}\label{sect_gener}%
When an EAS develops over the ocean surface Cherenkov radiation is generated not only in air. Some part of the cascade, depending on depth of the primary interaction as well as energy and direction of the shower, penetrates into water and also produces Cherenkov photons. Therefore we consider two media: air and water.

The discussion below will require direction cosines of the shower in the LRS def{}ined in section \ref{calc_tech}:
\begin{equation*}
\mu^\mathrm{sh}_x=-\sin\theta_\mathrm{sh}\cos\varphi_\mathrm{sh}\,,\quad\mu^\mathrm{sh}_y=-\sin\theta_\mathrm{sh}\sin\varphi_\mathrm{sh}\,,\quad
\mu^\mathrm{sh}_z=-\cos\theta_\mathrm{sh}\,,
\end{equation*}
where the zenith $\theta_\mathrm{sh}$ and azimuth $\varphi_\mathrm{sh}$ angles  are input parameters. Sought characteristics of a photon, namely its coordinates, direction cosines and weight, will be primed.

We assume that the photon is produced along the shower axis, consequently coordinates of its generation point can be calculated from
\begin{equation*}
z'=
\begin{cases}
-h_0\ln(X/X_0)\quad\quad\;\;\,\text{(air)},\\
\mu_z^\mathrm{sh}(X-X_0)/\rho_\mathrm{wat}\quad\text{(water)},
\end{cases}\quad
x'=z'\,\frac{\,\mu_x^\mathrm{sh}}{\,\mu_z^\mathrm{sh}}\,,\qquad
y'=z'\,\frac{\,\mu_y^\mathrm{sh}}{\,\mu_z^\mathrm{sh}}\,.
\end{equation*}
Here, $h_0\!=\!8.43\cdot10^5$\,cm is the standart atmosphere height, $X_0\!=\!1020/\cos\theta_\mathrm{sh}$ g/cm$^2$ is the ocean level depth, $\rho_\mathrm{wat}\!=\!1$\,g/cm$^3$ is the water density. The depth $X$ of the photon generation point is simulated from the Gaisser--Hillas formula (see, e.g.,\,\cite{Gaisser})
\begin{equation*}
N(X)=N_\mathrm{max}\biggl[\frac{X}{\,X_\mathrm{max}}\biggr]^{X_\mathrm{max}/\lambda}\exp\Bigl[(X_\mathrm{max}-X)/\lambda\Bigr]
\end{equation*}
with parameters describing the average cascade curve of a shower with primary energy $E_0\!=\!10^{20}$\,eV: $N_\mathrm{max}\!=\!7\!\cdot\!10^{10}$, $X_\mathrm{max}\!=\!800$\,g/cm$^2$, $\lambda\!=\!80$\,g/cm$^2$.

For f{}inding the photon direction in the LRS, it is convenient to introduce local reference systems associated with the shower axis (the reference system of the shower, RSS) and with the parent electron (the reference system of the electron, RSE). In the RSS the $Oz$--axis is steered along the shower axis and  angles $\theta_{e}$ and $\varphi_{e}$ characterize the direction of the electron, which, in turn, coincides with the $Oz$--axis of the RSE. The direction of the Cherenkov photon in the RSE is described by angles $\theta_\mathrm{ch}$ and $\varphi_\mathrm{ch}$. Then its direction cosines $\mu'_x$, $\mu'_y$ and $\mu'_z$ in the LRS are given by
\begin{equation*}
\begin{pmatrix}
\mu_x'\\
\mu_y'\\
\mu_z'
\end{pmatrix}\!=\!
\begin{pmatrix}
\gamma_\mathrm{sh}\,\mu^\mathrm{sh}_x\mu^\mathrm{sh}_z & -\gamma_\mathrm{sh}\,\mu^\mathrm{sh}_y & \mu^\mathrm{sh}_x\\
\gamma_\mathrm{sh}\,\mu^\mathrm{sh}_y\mu^\mathrm{sh}_z & \gamma_\mathrm{sh}\,\mu^\mathrm{sh}_x & \mu^\mathrm{sh}_y\\
-1/\gamma_\mathrm{sh} & 0 & \mu^\mathrm{sh}_z
\end{pmatrix}\!\times\!
\begin{pmatrix}
\gamma_e\,\mu^e_x\,\mu^e_z & -\gamma_e\,\mu^e_y & \mu^e_x\\
\gamma_e\,\mu^e_y\,\mu^e_z & \gamma_e\,\mu^e_x & \mu^e_y\\
-1/\gamma_e & 0 & \mu^e_z
\end{pmatrix}\!\times\!
\begin{pmatrix}
\mu_x^\mathrm{ch}\\
\mu_y^\mathrm{ch}\\
\mu_z^\mathrm{ch}
\end{pmatrix}\!\!,
\end{equation*}
where $\gamma_\alpha\!=\!\left[1-(\mu^\alpha_z)^2\right]^{-1/2}$ ($\alpha\!=\!\mathrm{sh},e$) and direction cosines of the electron and the photon are def{}ined by standart formulae:
\begin{equation*}
\mu_x^\beta=\sin\theta_\beta\cos\varphi_\beta\,,\quad
\mu_y^\beta=\sin\theta_\beta\sin\varphi_\beta\,,\quad
\mu_z^\beta=\cos\theta_\beta\quad(\beta=e,\mathrm{ch}).
\end{equation*}

In calculations we distinguish between ``forward--current'' ($\theta_e\!\leqslant\!80^\circ$) and ``back--current'' ($\theta_e\!>\!80^\circ$) electrons, with the latter taken into account only in generating in water. Zenith angles of the back--current electrons are distributed isotropically, while that of the forward--current electrons are simulated from the appropriate distribution suggested in\,\cite{Lafebre}. Azimuth angles of electrons of both types are uniformly distributed.

For photons, azimuth angles obey the uniform distribution and zenith angles  are given by the Tamm--Frank formula as a function of the electron energy $E$:
\begin{equation*}
\sin^2\!\theta_\mathrm{ch}=\frac{(E/E_\mathrm{th})^2-1}
{(E/m_ec^2)^2-1}\,.
\end{equation*}
Here, $E_\mathrm{th}\!=\!m_ec^2(1-n^{-2})^{-1/2}$ is the threshold generation energy of Cherenkov photons by an electron in a medium with refractive index $n$. To determine the energy of the forward--current electrons we use the suitable distribution from\,\cite{Lafebre}. The energy of the back--current electrons is found from the distribution
\begin{equation*}
\frac{\widetilde{N}(>E)}{N(>E_\mathrm{th})}=0.242
\left[3.1+\frac{E}{E_\mathrm{th}-m_ec^2}\right]^{-1.7},
\end{equation*}
which was specially derived by means of the Geant4 code\,\cite{Geant4} (the calculations were performed with a lower value of $E_0$). The number $\widetilde{N}$ of the back--current electrons with energies $E\!>\!E_\mathrm{th}$ is $\approx\!2.2$\,\% of~their total number $N$.

Finally, the photon is assigned an initial value of its weight. According to the Tamm--Frank theory, the number of photons with wavelength $\lambda$, $dq/d\lambda$, produced per unit depth, depends on the energy $E$ and depth $X$ of the parent electron and is given by
\begin{equation}\label{dq_dlambda}
\frac{\,dq}{\,d\lambda}(X,E)=\frac{2\pi\alpha}{\,\lambda^2\rho(X)}\sin^2\!\theta_\mathrm{ch}(E,E_\mathrm{th}(X)).
\end{equation}
Therefore the number of the photons in a ``packet'', the fate of which is simulated, varies from one iteration to another. We assume here that the unit--weight photon is generated by an electron with infinite energy at the ocean level. Then the weight at any generation point can be def{}ined as
\begin{equation*}
P^{\,\prime}=\frac{\,\rho_\mathrm{max}}{\rho}
\left[\frac{\sin\theta_\mathrm{ch}}{\,\sin\theta_\mathrm{ch}^{\,\mathrm{max}}}\right]^2\!.
\end{equation*}
We note that a value of $P^{\,\prime}$, averaged over a large number of events simulated, allows us to determine the total number of Cherenkov photons generated by electrons of an EAS quite easily (see app. \ref{app1}).

Thus, applying this scheme, we are able to specify the coordinates $x'$, $y'$, $z'$, direction cosines $\mu'_x$, $\mu'_y$, $\mu'_z$ and weight $P^{\,\prime}$ of a photon at its generation point, which are used further as input parameters in simulating its propagation in a medium.

\section{Scheme of photon propagation in a medium}\label{sect_prop}%
The brightest lines of f{}luorescent radiation in the atmosphere lie in the wavelength range $\lambda\!=\!300\!-\!400$\,nm. So the TUS detector is assumed to be sensitive to photons with these wavelengths and we, in turn, restrict our calculations within this range.

The process of photon propagation in a medium is usually represented as a sequence of scattering events, in which the coordinates of a new scattering point and direction of the photon are def{}ined. We consider scattering in water and air separately because of dif{}ferences in their physical properties. The input parameters are the coordinates, direction cosines and weight of the photon.

\subsection{Water}%
In simulating the photon travel we take scattering and absorption processes into account. The corresponding free path lengths $l_s$ and $l_a$ are given in\,\cite{Mobley}. The density and refractive index of water are assumed to be constant: $\rho_\mathrm{wat}\!=\!1$\,g/cm$^3$ and $n_\mathrm{wat}\!=\!1.33$.

Let us consider how a single scattering event is simulated. First, we calculate coordinates of the scattering point: $x_i\!=\!x'_i+\mu'_{x_i}l$ ($x_i\!=\!x,y,z$). Here, $l$ is the photon path length between successive scattering events, which, in the case of a medium with constant density, has the exponential distribution:
\begin{equation}\label{path_model}
p(l)\,dl=\exp(-l/l_s)\,dl/l_s\,.
\end{equation}
The weight of the photon is attenuated due to absorption on the path $l$:
$P=P^{\,\prime}\exp(-l/l_a)$. Then we compute new direction cosines of the photon in the LRS:
\begin{equation}\label{matr1}
\begin{aligned}
\begin{pmatrix}
\mu_x\\
\mu_y\\
\mu_z
\end{pmatrix}&\!=\!
\begin{pmatrix}
\gamma^{\,\prime}\mu'_x\mu'_z & -\gamma^{\,\prime}\mu'_y & \mu'_x\\
\gamma^{\,\prime}\mu'_y\mu'_z & \gamma^{\,\prime}\mu'_x & \mu^\prime_y\\
-1/\gamma^{\,\prime} & 0 & \mu'_z
\end{pmatrix}\!\!\times\!\!
\begin{pmatrix}
\sin\theta_s\cos\varphi_s\\
\sin\theta_s\sin\varphi_s\\
\cos\theta_s
\end{pmatrix}\quad\text{for}\;\;|\,\mu'_z|<1,\\
\begin{pmatrix}
\mu_x\\
\mu_y\\
\mu_z
\end{pmatrix}&\!=\,
\mathrm{sign}(\mu'_z)\!
\begin{pmatrix}
\sin\theta_s\cos\varphi_s\\
\sin\theta_s\sin\varphi_s\\
\cos\theta_s
\end{pmatrix}\quad\text{for}\;\;\left(1-|\,\mu'_z|\right)\leqslant10^{-5},
\end{aligned}
\end{equation}
where $\gamma^{\,\prime}\!=\!\left[1-\mu^{\prime\,2}_z\right]^{-1/2}$. The angles $\theta_s$ and $\varphi_s$ characterize def{}lection of the photon from its initial direction. $\varphi_s$ is uniformly distributed and $\theta_s$ is simulated from the Heneye--Greenstein distribution (see, e.g.,\,\cite{Mobley}):
\begin{equation*}
p(\theta_s,g)\,d\Omega_s=\frac1{4\pi}\,\frac{(1-g^2)\sin\theta_s}{(1+g^2-2g\cos\theta_s)^{3/2}}\,
d\theta_s\,d\varphi_s\,
\end{equation*}
with $g\!=\!0.924$.

Further the next scattering event is simulated. The procedure continues if the condition $P\!\geqslant\!P_\mathrm{min}$ holds or until the photon reaches the ocean surface.

\subsection{Air}%
Travelling of a photon in air is considered disregarding its ``true'' absorption. This assumption is quite acceptable in the wavelength range of photons under study here. Besides, we limit to Rayleigh scattering by presuming the absence of clouds, aerosols, etc. The atmosphere is supposed to be isothermal.

The coordinates, direction and weight of the photon are determined in the same way as for water. However the atmosphere is nonhomogeneous, which prohibits application of \eqref{path_model} for simulating the path length to the next scattering point. Nonetheless, within the assumptions indicated, the vertical depth $|\Delta X_v|$ traversed by the photon has the following simple exponential distribution (see app. \ref{app2}):
$$
p\,(\Delta X_v)\,d\Delta X_v=\frac{\,\varkappa}{\,\mu'_z}\,\exp(-\varkappa\,\Delta X_v/\mu'_z)\,d\Delta X_v\,.
$$
Hence we can f{}ind the vertical depth $X_v$ and coordinate $z$ of the scattering point:
$$
X_v=X'_v-\Delta X_v,\quad z=-h_0\ln(X_v/X_{v0}),
$$
where $X_{v0}\!=\!1020$\,g/cm$^2$ is the vertical depth at the sea level. The photon escapes the atmosphere boundary when $X_v\!<\!0$ and reaches the sea level when $X_v\!>\!1020$\,g/cm$^2$. In their turn, the path length and other coordinates of the scattering point are calculated as
$$
l=(z-z')/\mu'_z,\quad x=x'+\mu'_x\,l,\quad y=y'+\mu'_y\,l.
$$

The direction cosines of the photon in the LRS are determined from \eqref{matr1}, with $\varphi_s$ distributed uniformly and $\theta_s$ simulated from
$$
p(\theta_s)\,d\Omega_s=\frac{3}{\,16\pi}\,(1+\cos^2\theta_s)\sin\theta_s\,d\theta_s\,d\varphi_s\,.
$$

A new value of the photon weight is calculated with allowance for its attenuation due to scattering on the traversed path:
\begin{align*}
P&=P^{\,\prime}\frac{\;1-\exp(-\varkappa\,\Delta X_v)}{1-\exp(-\varkappa\,X'_v)}\quad\text{for}\;\;\mu'_z>0,\\
P&=P^{\,\prime}\frac{1-\exp(-\varkappa\,\Delta X_v)}{\,1-\exp\bigl[-\varkappa(X_{v0}-X'_v)\bigr]}\quad\text{for}\;\;\mu'_z<0,\\
P&=0\quad\text{for}\;\;\mu'_z\approx0.
\end{align*}
The weight does not change if the photon leaves the atmosphere boundary or reaches the ocean surface. This method of weight calculation makes it possible to ``select'' photons with low number of scattering events.

Further we simulate the next scattering event. The procedure continues if the condition $P\!\geqslant\!P_\mathrm{min}$ is true or until the photon reaches a required level.

\section{Wavy ocean surface}\label{sect_waves}%
We deal with two models of wavy ocean surface: with periodic structure (PS) and with chaotic structure (CS). In the f{}irst model the surface is represented in the LRS by a motionless wave with length $\lambda_w$ and amplitude $h_w$ along the $Ox$--axis. The shape of the wave is described by the parametric equations of a truncated cycloid:
\begin{equation*}
x=r_w\,t-h_w\sin t,\qquad z=h_w\cos t,
\end{equation*}
where $r_w\!=\!\lambda_w/2\pi$ and $r_w\!>\!h_w$, without dependence on the $y$--coordinate. In the second model the ocean surface is divided into many small f{}lat cells with height $\eta$. The quantities $\eta_u\!=\!\partial\eta/\partial x$ and $\eta_c\!=\!\partial\eta/\partial y$ (the subscripts $u$ and $c$ indicate the upwind and crosswind directions of the wave) follow the two-dimensional normal distribution \cite{Mobley}
\begin{equation}\label{distr_nu}
p\,(\eta_u, \eta_c)=\frac{1}{\,2\pi\sigma_u\sigma_c}\,\exp\!\left[-\frac{\,1}{\,2}\Bigl(\eta_u^2/\sigma_u^2+
\eta_c^2/\sigma_c^2\Bigr)\right]\!,
\end{equation}
with variances $\sigma_u^2\!=\!3.16\!\cdot\!10^{-3}u_w$, $\sigma_c^2\!=\!1.92\!\cdot\!10^{-3}u_w$ depending on the wind velocity $u_w$, in m/s. Both models assume that photons are ref{}lected and refracted at the interface according to the geometric optics laws. Let us consider the calculation procedure by the example of a ref{}lected photon.

{\it\bf Periodic structure of the ocean surface.} Initially we def{}ine the projections $x'$ and $z'$ of the coordinates $x$ and $z\!=\!0$ of~the~point at which the photon crosses the plane $Oxy$, onto the wavy surface with the assumption that one of the wave maxima coincides with the origin of the LRS. Hence the $x'$--coordinate is the residue of $x/\lambda_w$ and $z$ can be derived from the following equations: 
\begin{align*}
x'=r_w\arccos(z'/h_w)-\sqrt{h_w^2-z^{\prime\,2}}\quad\text{for}
\;\;x'\leqslant\lambda_w/2,\\
x'=2\pi r_w-r_w\arccos(z'/h_w)+\sqrt{h_w^2-z^{\prime\,2}}\quad\text{for}
\;\;x'>\lambda_w/2.
\end{align*}

Further we proceed to the reference system $Ox'y'z'$ in which the normal to the wavy surface at the point of the photon incidence serves as the $Oz'$--axis. Let $\mu_x$, $\mu_y$ and $\mu_z$ be initial direction cosines of the photon in the LRS. Then its direction cosines in the $Ox'y'z'$ system are def{}ined as 
\begin{equation*}
\begin{pmatrix}
\mu'_x\\
\mu'_y\\
\mu'_z
\end{pmatrix}\!\!=\!
\begin{pmatrix}
\cos\alpha\cos\beta & \sin\alpha\cos\beta & -\sin\beta\\
-\sin\alpha & \cos\alpha & 0\\
\cos\alpha\sin\beta & \sin\alpha\sin\beta & \cos\beta
\end{pmatrix}\!\!\times\!\!
\begin{pmatrix}
\mu_x\\
\mu_y\\
\mu_z
\end{pmatrix}\!\!,
\end{equation*}
where $\alpha\!=\!0$ for $x'\!\leqslant\!\lambda_w/2$ and $\alpha=\pi$ for $x'\!>\!\lambda_w/2$, and $\beta$ is the angle between axes $Oz'$ and $Oz$:
$$
\beta=\arccos\frac{r_w-z'}{\,\sqrt{h_w^2+r_w^2-2r_wz'\,}\,}\,.
$$

Finally we determine the direction cosines $\nu_x^{\,\prime}$, $\nu_y^{\,\prime}$, $\nu_z^{\,\prime}$ of the photon ref{}lected and appropriate ref{}lection coef{}f{}icient. Transformation to the LRS is performed from
\begin{equation*}
\begin{pmatrix}
\nu_x\\
\nu_y\\
\nu_z
\end{pmatrix}\!\!=\!
\begin{pmatrix}
\cos\alpha\cos\beta & -\sin\alpha & \cos\alpha\sin\beta \\
\sin\alpha\cos\beta & \cos\alpha & \sin\alpha\sin\beta\\
-\sin\beta & 0 & \cos\beta
\end{pmatrix}\!\!\times\!\!
\begin{pmatrix}
\nu_x^{\,\prime}\\
\nu_y^{\,\prime}\\
\nu_z^{\,\prime}
\end{pmatrix}\!\!.
\end{equation*}

{\it\bf Chaotic structure of the ocean surface.}
At the first step we simulate the tangents $\eta_u$ and $\eta_c$ of the angles between a cell and $Ox$ and $Oy$ axes, respectively, from \eqref{distr_nu} and determine the direction cosines of the normal in the LRS at each point of the photon incidence:
$$
\zeta_x=\frac{\eta_u}{\sqrt{\eta^2_u+\eta^2_c+1}\,}\,,\quad
\zeta_y=\frac{\eta_c}{\sqrt{\eta^2_u+\eta^2_c+1}\,}\,,\quad
\zeta_z=\frac1{\sqrt{\eta^2_u+\eta^2_c+1}}\,.
$$

Further the direction cosines of the incident photon in the system of the normal are calculated:
\begin{equation}\label{wave_matr}
\begin{pmatrix}
\mu'_x\\
\mu'_y\\
\mu'_z
\end{pmatrix}\!\!=\!
\begin{pmatrix}
\gamma\,\zeta_x\zeta_z & \gamma\,\zeta_y\zeta_z & -1/\gamma \\
-\gamma\,\zeta_y  & \gamma\,\zeta_x & 0\\
\zeta_x & \zeta_y & \zeta_z
\end{pmatrix}\!\!\times\!\!
\begin{pmatrix}
\mu_x\\
\mu_y\\
\mu_z
\end{pmatrix}\!\!,
\end{equation}
where $\gamma\!=\!(1-\zeta^2_z)^{-1/2}$. The direction cosines of the ref{}lected photon and its ref{}lection coef{}f{}icient are found from the laws of geometric optics.

Finally transformation to the LRS is implemented by means of the rotation matrix inverse to the matrix from \eqref{wave_matr}.

\section{Recording Cherenkov light signal by the TUS detector}%
The simulation algorithm presented above was applied in \cite{Shustova} for estimating the signal from ref{}lected EAS Cherenkov light by the TUS detector with field of view of $\approx\!9.5^\circ$. Within the model of periodic structure of a wave, with amplitude $h_w\!=\!0.7$\,m and length $\lambda_w\!=\!40$\,m, the registration of f{}lashes of the ref{}lected photons was found to be possible for showers with zenith angles $\lesssim\!20^\circ$. The wave parameters indicated were chosen on the basis of the following data. Satellite observations \cite{wind_speed} show that, on average, winds under 8\,m/s in speeds predominate over the ocean. According to the Beaufort scale, they are matched to waves with heights up to 2\,m. It is known that there exists a wavelength spectrum at a given amplitude $h_w$. Using a two--dimensional distribution over half--lengths and amplitudes, suggested in \cite{wave_parameters}, we chose the mean wavelength, $\lambda_w\!=\!0.7$\,m, which corresponds to $h_w\!=\!0.7$\,m.

In this section we compare the calculation results, obtained within both models. Note that the CS model seems to be more realistic as constructing a kind of ``ripple'' on the ocean surface. Moreover, it depends on only one parameter, the wind velocity. However the situation where a grid of chaotic ``f{}lutters'' is superimposed on an ordered structure, appeared to us to be more appropriate in the case of intense waves.

\begin{figure}[h!]
\begin{center}
\includegraphics[height=16pc]{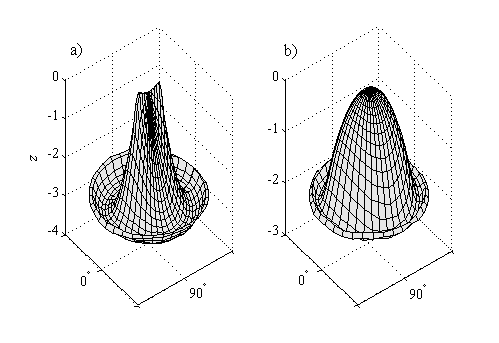}
\end{center}
\vspace{-1.5pc}
\caption{Angular distribution of ref{}lected Cherenkov photons near the ocean surface for showers with angles $\theta_\mathrm{sh}\!=\!0^\circ$ and $\varphi_\mathrm{sh}\!=\!0^\circ$, obtained within the models of a) periodic and b) chaotic structures of the ocean surface. The $Oz$--axis specifies the relative density of the photons on logarithmic scale.}\label{pic1}
\end{figure}

Let us adopt that the wave parameters $h_w\!=\!0.7$\,m and $\lambda\!=\!40$\,m in the PS model are matched to the wind velocity $u_w\!=\!6$\,m/s in the CS model. Fig.\,\ref{pic1} presents angular distributions of ref{}lected Cherenkov photons near the ocean surface for vertical showers. Photons with zenith $\theta$ and azimuth $\varphi$ angles impinge on proper cells with area $r\Delta r\Delta\varphi$, where $\Delta r\!=\!0.05$ and $\Delta\varphi\!=\!\pi/18$. A value of $r$ varies from 0 to 1 and specif{}ies $\sin\theta$. The $z$--coordinate specifies the decimal logarithm of the photon relative surface density. As illustrated, values of $\theta$ in the CS model lies in a much wider range.
\vspace{-0.5pc}

\begin{figure}[h!]
\begin{center}
\includegraphics[height=16pc]{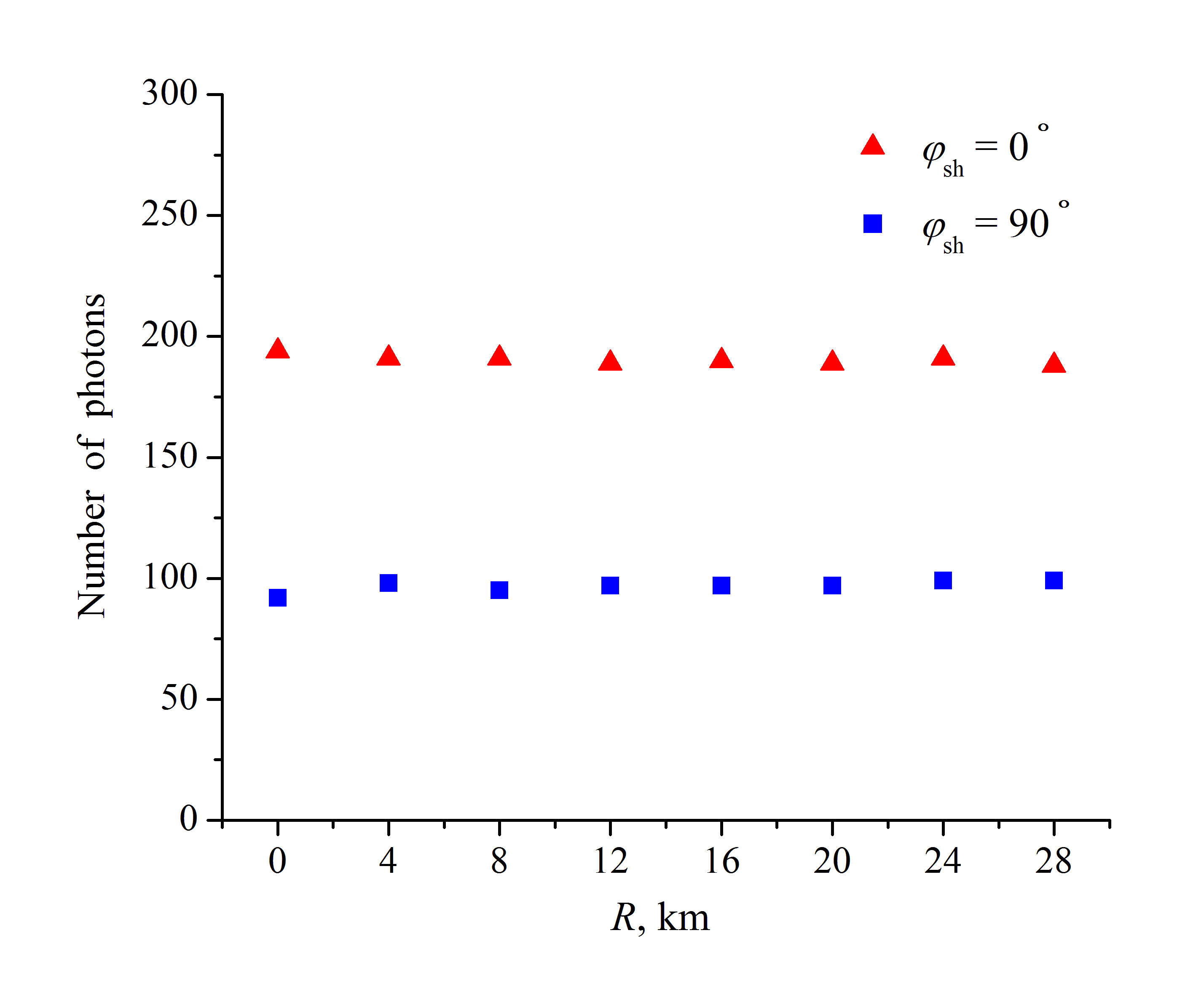}
\end{center}
\vspace{-1.5pc}
\caption{Number of ref{}lected Cherenkov photons that hit the detector mirror, as a function of its location $R$, in~the CS model for showers with angles $\theta_\mathrm{sh}\!=\!25^\circ$, $\varphi_\mathrm{sh}\!=\!0^\circ$ and $\varphi_\mathrm{sh}\!=\!90^\circ$. The point $R\!=\!0$ at the height of 400\,km is projected onto the center of the LRS.}\label{pic2}
\end{figure}

The restriction on EAS zenith angles, $\theta_\mathrm{sh}\!\lesssim\!20^\circ$, has been obtained within the PS model with allowance to the condition $\varphi_\mathrm{sh}\!=\!90^\circ$, when the shower axis is in one plane with the wavefront and therefore more photons are ref{}lected from a practically flat surface. This is also true for the CS model because $\sigma_c\!<\!\sigma_u$. The azimuthally average number of the photons that will hit the mirror of the TUS detector is shown in Fig.\,\ref{pic2} for showers with $\theta_\mathrm{sh}\!=\!25^\circ$, $\varphi_\mathrm{sh}\!=\!0^\circ$ and $\varphi_\mathrm{sh}\!=\!90^\circ$.

\begin{table}[h!]
\begin{center}
\begin{tabular}{|c|c|c|c|c|}
\hline
$R$, &\multicolumn{4}{|c|}{$u_w$, m/s}\\
\cline{2-5}
km & 2 & 4 & 6 & 8\\
\hline
$\vphantom{\Bigl|}$0 & $17.7\pm5.0$ & $58.5\pm5.1$ & $92.0\pm5.0$ & $107.6\pm5.0$\\
4 & $21.5\pm0.6$ & $62.2\pm0.7$ & $97.9\pm0.7$ & $112.6\pm0.7$\\
$\vphantom{\Bigl|}$8 & $21.6\pm0.3$ & $62.8\pm0.4$ & $95.3\pm0.4$ & $113.5\pm0.4$\\
12 & $22.2\pm0.2$ & $63.3\pm0.2$ & $96.1\pm0.3$ & $113.7\pm0.3$\\
$\vphantom{\Bigl|}$16 & $23.0\pm0.2$ & $65.3\pm0.2$ & $96.5\pm0.2$ & $114.0\pm0.2$\\
20 & $24.2\pm0.1$ & $66.2\pm0.2$ & $97.1\pm0.2$ & $114.1\pm0.2$\\
$\vphantom{\Bigl|}$24 & $25.3\pm0.1$ & $68.0\pm0.1$ & $99.4\pm0.2$ & $115.3\pm0.2$\\
28 & $27.2\pm0.1$ & $69.2\pm0.1$ & $99.3\pm0.1$ & $115.1\pm0.1$\\
\hline
\end{tabular}
\end{center}
\caption{Number of reflected Cherenkov photons that impinge upon the mirror of the TUS detector in the CS model for showers with angles $\theta_\mathrm{sh}\!=\!25^\circ$ and $\varphi_\mathrm{sh}\!=\!90^\circ$.}
\label{tab1}
\end{table}

Time resolution of the TUS detector ($\approx\!1\mu$s) should be taken into account in order to estimate the observed Cherenkov signal value. For small zenith angles of showers, the photons can be considered to arrive at the detector mirror simultaneously. However, the signal duration grows starting from $\theta_\mathrm{sh}\!=\!25^\circ\!-\!30^\circ$ and therefore the signal amplitude reduces even despite of the fact that photons grow in number with an increase of the shower zenith angle. Our calculations show that in the CS model, with $u_w\!=\!6$\,m/s, f{}lashes of ref{}lected Cherenkov photons at the end of the f{}luorescence track can be recorded by the TUS detector provided that $\theta_\mathrm{sh}\!\lesssim\!25^\circ$.

With a change in wind velocity the restriction on $\theta_\mathrm{sh}$ will also change because of proportionality of the variances $\sigma^2_u$ and $\sigma^2_c$ to the wind velocity. In tab.\,\ref{tab1} are listed the results of the calculations, obtained within the CS model at different values of $u_w$ for showers propagating at angles $\theta_\mathrm{sh}\!=\!25^\circ$ and $\varphi\!=\!90^\circ$. The entries of the columns indicate the number of ref{}lected Cherenkov photons that fall within the field of view of the TUS detector, as a function of $R$. As evident, the photons are quite small in number at $u_w\!=\!2$\,m/s so the maximum value of the shower zenith angle should be decreased.

\section*{Conclusions}%

In this paper we have presented the simulation algorithm for generating and propagating EAS Cherenkov photons over the ocean. We have considered two dif{}ferent models for describing the wavy ocean surface and imposed restrictions on zenith angles of the showers, at which the TUS detector would be able to identify the f{}lashes of ref{}lected Cherenkov photons. Our calculations have shown that despite the dif{}ference in number of the photons impinging on the detector, both models provide approximately the same results.

\section*{Acknowledgments}%

This work was supported by Russian Foundation for Basic Research (grant 09-02-12162-ofi\_m).

\appendix
\section{Number of Cherenkov photons in a shower}\label{app1}%

The number of photons with wavelength $\lambda$, produced in a layer $(X_2\!-\!X_1)$, is written as
\begin{equation}\label{photon_num}
\frac{\,dN_\gamma}{d\lambda}=\!\!\int\limits_{X_1}^{\;X_2}\!\!dX'\,N(X')
\!\!\!\!\int\limits_{E_\mathrm{th}(X)}^{\;\infty}\!\!\!\!dE'\frac{\,dN}{\,dE'}(X',E')\,\frac{\,dq}{\,d\lambda}(X',E').
\end{equation}
The distributions for Cherenkov photons, $dq/d\lambda$, and for electrons, $N(x)$ and $dN(X,E)/dE$, are introduced in section \ref{sect_gener}.

We recall that every electron is assigned not a packet but rather a single photon with weight
\begin{equation*}
P(X,E)=\frac{\,\rho_\mathrm{max}}{\rho}
\left[\frac{\sin\theta_\mathrm{ch}}{\,\sin\theta_\mathrm{ch}^\mathrm{\,max}}\right]^2\!.
\end{equation*}
Therefore, \eqref{photon_num} can be rewritten as
\begin{align*}
\frac{\,dN_\gamma}{d\lambda}=
\frac{\,2\pi\alpha}{\lambda^2}\,\frac{\,\sin^2\!\theta_\mathrm{ch}^\mathrm{\,max}}
{\rho_\mathrm{max}}\!\int\limits_{X_1}^{\;X_2}\!\!dX'\,N(X')
\!\!\!\!\int\limits_{E_\mathrm{th}(X')}^{\;\infty}\!\!\!\!dE'\,&\frac{\,dN}{\,dE'}(X',E')\,P(X',E')\approx\\
&\approx\widetilde{P}\,\frac{\,2\pi\alpha}{\lambda^2}\,\frac{\,\sin^2\!\theta_\mathrm{ch}^\mathrm{\,max}}
{\rho_\mathrm{max}}\!\int\limits_{X_1}^{\;X_2}\!\!dX'\,N(X).
\end{align*}
The approximation sign takes in account the fact that $\Bigl(\,\sum_{i=1}^N P_{i}\Bigr)/N$ tends to $\widetilde{P}$ with an increase in the number $N$ of events simulated. 

\section{Distribution for vertical depth}\label{app2}%
Let us consider a change in intensity of a light beam on a path length $dl$:
\begin{equation}\label{dI}
dI=-I\,dl/l_s\,.
\end{equation}
For Rayleigh scattering, the mean free path length $l_s$ depends on the wavelength $\lambda$ of radiation and density $\rho$ of the medium:
\begin{equation*}
l_s\propto\frac{\,\lambda^4}{\rho}\,.
\end{equation*}

Introducing the quantity $\varkappa\!=\!(l_s\rho)^{-1}$ and using the following relationship between the vertical depth $X_v$ of the atmosphere and the height $h$ above the sea level:
$$
X_v(h)=\!\int\limits^{\;\infty}_{\!\!h}\!\!\rho(h')\,dh',
$$
we integrate \eqref{dI} over a layer with thickness $(h_2-h_1)$:
\begin{equation}\label{intensity}
\ln\!\left[\frac{\,I(h_2)}{I(h_1)}\right]\!=-\varkappa\!\int\limits_{h_1}^{\;\;h_2}\!\!\rho(h')\,dh'=-\varkappa\Bigl[X_v(h_1)-X_v(h_2)\Bigr]=
-\varkappa\,\Delta X_v.
\end{equation}
If the beam is def{}lected at an angle $\theta$ from the vertical direction, \eqref{intensity} transforms to
\begin{equation*}
\ln\!\left[\frac{I(l_2)}{I(l_1)}\right]\!=
-\varkappa\!\int\limits_{l_1}^{\;\;l_2}\!\!\rho(l')\,dl'=
-\varkappa\Bigl[X_v(h_1)-X_v(h_2)\Bigr]/\cos\theta=-\varkappa\,\Delta X_v/\mu_z\,,
\end{equation*}
where $(l_2-l_1)\!=\!(h_2-h_1)/\mu_z$. We implicitly assume here the case of the plane-parallel atmosphere, which is in reality valid up to $\theta\!\approx\!70^\circ$. In our calculations photons with $\mu_z\!\approx\!0$ are removed by setting their weight to zero.

Thus the probability that a photon traverses the vertical depth $|\Delta X_v|$ without scattering obeys the following exponential law:
$$
p\,(\Delta X_v)\,d\Delta X_v=\frac{\varkappa}{\,\mu_z}\,\exp(-\varkappa\,\Delta X_v/\mu_z)\,d\Delta X_v\,.
$$


\begin{thebibliography}{99}

\bibitem{tus}
\arten{B.A.\,Khrenov, V.V.\,Alexandrov, D.I.\,Bugrov et al.}{Phys. Atom. Nucl.}{67}{2058}{2004}

\bibitem{balloon}
\arten{S.B.\,Shaulov, S.P.\,Besshapov, N.V.\,Kabanova et al.}{Nucl. Phys., B, Proc. Suppl.}{196}{403}{2009}

\bibitem{Antonov}
{\it R.A.\,Antonov, D.V.\,Chernov, M.\,Finger et al.,} Proc. 28th ICRC. Tsukuba. 981 (2003).

\bibitem{Como}
\arten{O.P.\,Shustova, N.N.\,Kalmykov, B.A.\,Khrenov}{Proc. 12th ICATPP, Astropart. Part. Space Phys. Rad. Int. Med. Phys.}{6}{290}{2011}

\bibitem{Shustova}
\artenarx{O.P.\,Shustova, N.N.\,Kalmykov, B.A.\,Khrenov}{Bull. Rus. Acad. Sci. Phys.}{75}{381}{2011}{arXiv:1110.2974}

\bibitem{jem_euso}
\artenarx{Y.\,Takahashi and the JEM-EUSO Collaboration}{New Jour. Phys.}{11}{065009}{2009}{arXiv:0910.4187}

\bibitem{Gaisser}
{\it T.K.\,Gaisser,} Cosmic Rays and Particle Physics. Cambridge: Cambridge University Press, 1990.

\bibitem{Lafebre}
\artenarx{S.\,Lafebre, R.\,Engel, H.\,Falcke et al.}{Astropart. Phys.}{31}{243}{2009}{arXiv:0902.0548}

\bibitem{Geant4}
{\it The Geant4 Collaboration}, Physics Reference Manual. Version: geant4.9.3 (2009)\\\bibliolink{http://geant4.web.cern.ch}.

\bibitem{Mobley}
{\it C.D.\,Mobley,} Light and Water: Radiative Transfer in Natural Waters. London: Academic Press, 1994.

\bibitem{wind_speed}
{\it Winds. Measuring Ocean Winds from Space.} \bibliolink{http://winds.jpl.nasa.gov}

\bibitem{wave_parameters}
\arten{J.\,Ryden, S.\,van\,Iseghem, M.\,Olagnon, I.\,Rychlik}{Appl. Ocean Res.}{24}{189}{2002}


\end{thebibliography}
\end{document}